\documentclass[superscriptaddress,twocolumn,showpacs,aps,prl]{revtex4-1}
\usepackage{graphicx}
\usepackage{amsmath}
\usepackage[colorlinks=true]{hyperref}
\DeclareMathOperator{\sech}{sech}
\DeclareMathOperator{\csch}{csch}
\DeclareMathOperator{\sgn}{sgn}

\begin{document}
\title{Annihilation of Domain Walls in a Ferromagnetic Wire}
\author{Anirban Ghosh}
\affiliation{Department of Physics and Astronomy, Johns Hopkins University, Baltimore, Maryland 21218, USA}
\author{Kevin S. Huang}
\affiliation{Department of Physics and Astronomy, Johns Hopkins University, Baltimore, Maryland 21218, USA}
\affiliation{Centennial High School, Ellicott City, Maryland 21042, USA}
\author{Oleg Tchernyshyov}
\affiliation{Department of Physics and Astronomy, Johns Hopkins University, Baltimore, Maryland 21218, USA}

\begin{abstract}
We study the annihilation of topological solitons in the simplest setting: a one-dimensional ferromagnet with an easy axis. We develop an effective theory of the annihilation process in terms of four collective coordinates: two zero modes of the translational and rotational symmetries $Z$ and $\Phi$, representing the average position and azimuthal angle of the two solitons, and two conserved momenta $\zeta$ and $\varphi$, representing the relative distance and twist. Comparison with micromagnetic simulations shows that our approach captures well the essential physics of the process.
\end{abstract}

\maketitle

The dynamics of topological solitons in ferromagnets \cite{Kosevich1990} poses a class of problems of fundamental interest. Time evolution of magnetization is governed by the Landau-Lifshitz-Gilbert (LLG) equation \cite{Landau1935, Gilbert2004}
\begin{equation}
\mathcal{J}\dot{\mathbf{m}}=\mathbf{m}\times\mathbf{h}_{\mathrm{eff}}+\alpha|\mathcal{J}|\dot{\mathbf{m}}\times\mathbf{m}
\label{eq:LLG}
\end{equation}
Here $\mathbf{m}(\mathbf r, t) = \mathbf{M}/\mathbf{|M|}$ is the unit-vector field of magnetization, $\mathcal{J}=|\mathbf{M}|/\gamma$ is the angular momentum density, $\mathbf{h}_{\mathrm{eff}}(\mathbf{r})=-\delta U/\delta\mathbf{m}(\mathbf{r})$ is the effective magnetic field obtained from the potential energy functional $U[\mathbf{m}(\mathbf{r})]$ and $\alpha\ll 1$ is the Gilbert damping constant. Since the magnetization field has infinitely many modes that are coupled non-linearly, full analytic solution to a dynamical problem is unavailable in most cases. 

\begin{figure}
\includegraphics[width=0.95\columnwidth]{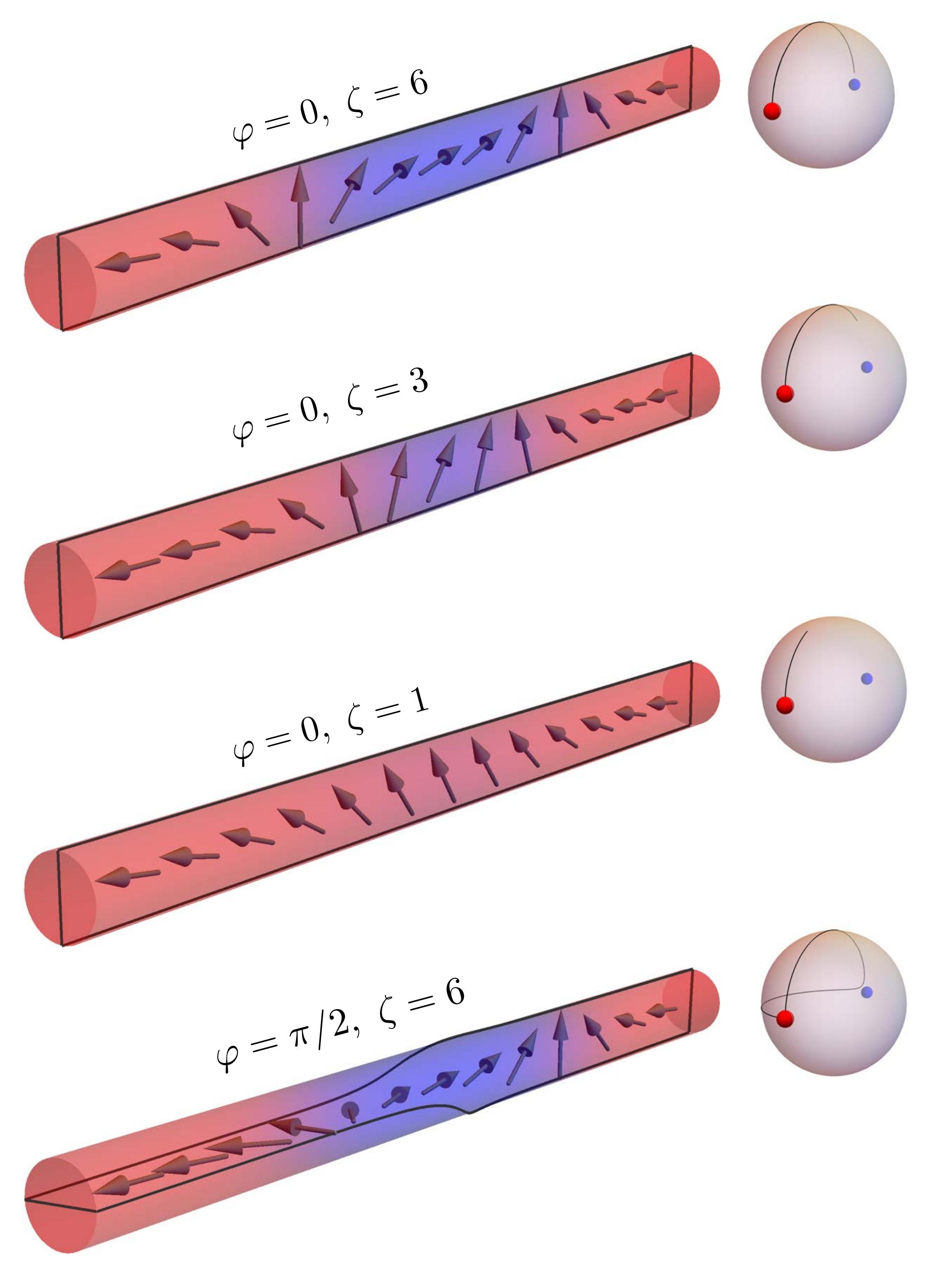}
\caption{(Color online) Several configurations of a pair of domain walls with shown values of separation $\zeta$ and twist $\varphi$. The red and blue colors denote positive and negative magnetization component $m_z$ along the axis of the cylinder. The wire frames depict the local plane tangential to the magnetization field. Spheres on the right show the path of the magnetization field $\mathbf m(z)$ as $z$ goes from $-\infty$ to $+\infty$, beginning from and ending at the north pole (red). The south pole (blue) can only be reached if the separation of the domain walls $\zeta = \infty$.}
\label{fig:soliton}
\end{figure}

A powerful alternative approach is to identify a small number of relevant soft modes, parametrized in terms of collective coordinates, and formulate an effective theory only in terms of these. This method was first applied to magnetic solitons by \textcite{Walker1973} to describe the dynamics of a domain wall in an easy-axis ferromagnet in one spatial dimension, $\mathbf{m}=\mathbf{m}(z,t)$, with the Lagrangian \cite{Kosevich1990}
\begin{equation}
L = \int_{-\infty}^\infty dz \, 
		\mathcal{J}(\cos\theta-1)\, \dot{\phi} - U,
\label{eq:L}
\end{equation}
and the potential energy
\begin{equation}
U = \int_{-\infty}^\infty dz 
	\left(
		A |\mathbf m'|^2 + K |\mathbf m\times \hat{\mathbf z}|^2 
	\right)/2.
\label{eq:U}
\end{equation}
Here $\theta$ and $\phi$ are the polar and azimuthal angles of magnetization $\mathbf m$, $A$ is the exchange constant, $K$ is the anisotropy, and $\hat{\mathbf z}$ is the direction of the easy axis. The unit of length is the width of the domain wall $\ell_0 = \sqrt{A/K}$ and the unit of time is the inverse of the ferromagnetic resonance frequency, $t_0 = 1/\omega_0 = \mathcal J/K$. In what follows, we work in these natural units and set $\mathcal{J} = A = K = \ell_0 = t_0 = 1$. A topological soliton interpolating between the two ground states $\mathbf m = \pm \hat{\mathbf z}$ and minimizing the potential energy (\ref{eq:U}) is a domain wall
\begin{equation}
\cos{\theta(z)} = \pm \tanh{(z-Z)}, 
\quad
\phi(z) = \Phi.
\label{eq:domain-wall}
\end{equation}
The position of a domain wall $Z$ and its azimuthal angle $\Phi$ are collective coordinates describing the zero modes associated with the translational and rotational symmetries. Schryer and Walker showed that, in the presence of weak perturbations, the dynamics of a domain wall reduces to a time evolution of $Z$ and $\Phi$. By substituting the domain-wall Ansatz (\ref{eq:domain-wall}) into the LLG equation (\ref{eq:LLG}) or into the Lagrangian (\ref{eq:L}), one obtains an effective theory for this system in terms of the two collective coordinates $\Phi$ and $Z$ \cite{Walker1973}. Their equations of motion read 
\begin{subequations}
\begin{eqnarray}
F_\Phi \pm 2 \dot{Z} - 2 \alpha \dot{\Phi} = 0,
\\
F_Z \mp 2 \dot{\Phi} - 2 \alpha \dot{Z} = 0,
\label{eq:dynamics-domain-wall}
\end{eqnarray}
\end{subequations}
where the force $F_Z = - \partial U/\partial Z$ and the torque $F_\Phi = - \partial U/\partial \Phi$ are derived from potential energy $U$ that may include perturbations beyond Eq.~\ref{eq:U}. 

More generally, a magnetic soliton can be described by a set of time-dependent collective coordinates $\mathbf q(t) = \{q_1(t), q_2(t), \ldots\}$, whose equations of motion express the balance of conservative, gyrotropic, and viscous forces for each coordinate $q_i$ \cite{Tretiakov2008}:
\begin{subequations}
\begin{eqnarray}
F_i &+& G_{ij}\dot{q}_j - \Gamma_{ij}\dot{q}_j = 0,
\label{eq:q-equations-eqn}
\\
F_i &=& - \frac{\partial U}{\partial q_i},
\label{eq:q-equations-F}
\\
G_{ij} &=& -\int dV\, 
	\mathbf{m} \cdot 
	\left( 
		\frac{\partial\mathbf{m}}{\partial q_{i}} 
		\times \frac{\partial\mathbf{m}}{\partial q_{j}} 
	\right),
\label{eq:q-equations-G}
\\
\Gamma_{ij} &=& \alpha \int dV\, 
	\frac{\partial\mathbf{m}}{\partial q_{i}} 
	\cdot \frac{\partial\mathbf{m}}{\partial q_{j}}.
\label{eq:q-equations-Gamma}
\end{eqnarray}
Here $F_i$ is the conservative force conjugate to collective coordinate $q_i$, $G_{ij}$ is the antisymmetric gyrotropic tensor, and $\Gamma_{ij}$ is the symmetric viscosity tensor.
\label{eq:q-equations}
\end{subequations}

The method of collective coordinates has been quite successful in describing the dynamics of solitons in ferromagnets \cite{Ivanov1989, Guslienko2006, Wong2010, Makhfudz2012, Kim2015} and antiferromagnets \cite{Ivanov1995, Tveten2013, Kim2014, Tveten2016}. In most cases, the set of coordinates $\mathbf q$ is limited to just the zero modes associated with the global symmetries of the unperturbed system. In such a case, weak perturbations create a gentle potential landscape $U(\mathbf q)$ that induces slow dynamics of the formerly zero and now soft modes $\mathbf q$, while the hard modes quickly adjust to the instantaneous configuration of the soft modes. Including hard modes as dynamical degrees of freedom poses significant challenges \cite{Clarke2008}.

Here we apply the method of collective coordinates to a problem that requires going beyond the zero-mode approximation: the annihilation of two domain walls in a one-dimensional ferromagnet with an easy axis. When two domain walls are far apart, they behave like rigid objects and can be described by two independent pairs of collective coordinates: two positions $Z_{1}$ and $Z_{2}$ and two azimuthal angles $\Phi_{1}$ and $\Phi_{2}$.  Alternatively, we may use the average position $Z=(Z_1+Z_2)/2$ and the average angle $\Phi=(\Phi_1+\Phi_2)/2$ and two relative coordinates, the separation $\zeta = Z_2 - Z_1$ and the twist $\varphi= \Phi_2-\Phi_1$. Whereas $Z$ and $\Phi$ represent the zero modes associated with the symmetries of translation and rotation, the relative coordinates $\zeta$ and $\varphi$ affect the energy (\ref{eq:U}) and thus represent modes that harden as the domain walls get closer and their interaction increases. 

Let us work with the boundary condition $\mathbf{m}(\pm\infty)=\hat{\mathbf{z}}$. We may anticipate how the annihilation proceeds in the limit of large separation, $\zeta \gg 1$, when the domain walls retain their individual character and are described by Eq.~\ref{eq:dynamics-domain-wall} with the top signs. The effects on the average and relative coordinates occur at different orders in $\alpha$. To zeroth order, the two domain walls exhibit rigid rotational and translational motion:
\begin{equation}
\dot{\Phi} = - \frac{1}{2} \frac{\partial U}{\partial \zeta},
\quad
\dot{Z} = \frac{1}{2} \frac{\partial U}{\partial \varphi}.
\end{equation}
They acquire relative velocities at the next order:
\begin{equation}
\dot{\varphi} = - \alpha \frac{\partial U}{\partial \varphi},
\quad
\dot{\zeta} = - \alpha \frac{\partial U}{\partial \zeta}.
\end{equation}

As the domain walls approach each other and begin to overlap, they lose their ideal shape (\ref{eq:domain-wall}) and Eq.~\ref{eq:dynamics-domain-wall} no longer applies. Even worse, 
the precise positions and azimuthal angles of overlapping domain walls become ill-defined. Fortunately, we may use two conserved momenta---angular $J$ and linear $P$---as proxies for the separation and twist. The angular momentum along the $z$ axis is \cite{Kosevich1990}
\begin{equation}
J = \int_{-\infty}^\infty dz \, (\cos{\theta}-1). 
\label{eq:J}
\end{equation}
Here the subtraction of 1 in the brackets means that we measure the angular momentum relative to the uniform ground state $\mathbf m = \hat{\mathbf z}$. If the domain walls are far apart, $\zeta \gg 1$, $\cos{\theta} \approx -1$ in the space between them, so $J \approx -2 \zeta$. Turning this around, we \emph{define} the separation in terms of the angular momentum (\ref{eq:J}), $\zeta \equiv -J/2$. 

The problem with the relative twist is fixed in a similar way. The linear momentum of a non-topological soliton (i.e., one approaching the same ground state at both ends, $\mathbf m(\pm \infty) = \hat{\mathbf z}$) is \cite{Kosevich1990}
\begin{equation}
P = \int_{-\infty}^\infty dz \, (1 - \cos{\theta}) \phi' 
	= \oint (1 - \cos{\theta}) d\phi.
\label{eq:P}
\end{equation}
The linear momentum is the area subtended by the vector $\mathbf m(z)$ on the unit sphere as $z$ goes from $-\infty$ to $+\infty$ \cite{Haldane1986}. For two well-separated domain walls with a twist $\varphi$, this area is $2\varphi$. Again, we turn things around and \emph{define} the twist in terms of the linear momentum (\ref{eq:P}), $\varphi \equiv P/2$. Pairs of domain walls with several values of separation $\zeta$ and twist $\varphi$ are shown in Fig.~\ref{fig:soliton}.

Unlike single domain walls, which are stable for topological reasons, pairs of domain walls are unstable: minimization of the energy (\ref{eq:U}) in the topologically trivial sector with $\mathbf m(\pm \infty) = \hat{\mathbf z}$ yields a uniform ground state $\mathbf m(z) = \hat{\mathbf z}$. To obtain a solution for a pair of domain walls, we may rely on conservation of linear and angular momenta and minimize the energy $U$ at fixed $P$ and $J$. This can be done through minimization of the modified energy 
\begin{equation}
\tilde{U} = U - PV - J \Omega,
\label{eq:U-tilde}
\end{equation}
where $V$ and $\Omega$ are Lagrange multipliers. The corresponding Lagrangian, 
\begin{equation}
\tilde{L} = \int_{-\infty}^\infty dz \, 
		(\cos{\theta-1})(\dot{\phi} - V \phi' + \Omega) - U,
\end{equation}
describes the dynamics of magnetization in a new frame moving at the linear velocity $V$ and rotating at the angular velocity $\Omega$. Minimization of the new potential energy (\ref{eq:U-tilde}) yields a static soliton in the new frame. In the static frame, the soliton is moving at the velocity $V$ and is rigidly rotating at the frequency $\Omega$. This class of dynamic solitons was first obtained by \textcite{Kosevich1977} and by \textcite{Long1979}. The relation between the velocities and momenta of these solutions is
\begin{equation}
V = - \frac{2\sin{2P}}{\sinh{2J}},
\quad
\Omega = \frac{\sin^2{P}}{\sinh^2{J}} - \frac{\cos^2{P}}{\cosh^2{J}}.
\label{eq:velocities-momenta}
\end{equation}
The explicit form of the solitons is given in the Supplemental Material \cite{supplemental}. 

We are now ready to derive the equations of motion for a pair of domain walls with four collective coordinates $\Phi$, $Z$, $\varphi$, and $\zeta$ by using the general formalism (\ref{eq:q-equations}). The gyrotropic coefficients are most easily derived from the Berry phase term in the effective Lagrangian for the collective coordinates. They form two pairs of conjugate variables, $\Phi$ and $J = -2\zeta$ for rotational motion and $Z$ and $P = 2\varphi$ for translational. We thus infer that the effective Lagrangian includes the Berry-phase terms \cite{Clarke2008}
\begin{equation}
L_B = -2\zeta\dot{\Phi}+2\varphi\dot{Z} = A_i \dot{q}_i.
\end{equation}
From that we read off the Berry connections $A_{\Phi} = -2\zeta$, $A_{Z} = 2\varphi$, and $A_{\zeta} = A_{\varphi} = 0$. The gyrotropic coefficients are the Berry curvatures $G_{ij} = \partial_{i}A_{j}-\partial_{j}A_{i}$. The nonzero coefficients of the gyrotropic tensor are 
\begin{equation}
G_{\Phi \zeta} = - G_{\zeta \Phi} 
	= G_{\varphi Z} = -G_{Z \varphi} = 2.
\label{eq:G-nonzero}
\end{equation}

To deduce conservative forces $F_i$, we turn off dissipation. Eq.~\ref{eq:q-equations} now read $F_i + G_{ij} \dot{q}_j = 0$. Conservation of linear and angular momenta implies the absence of the external force and torque, $F_Z = F_\Phi = 0$. The relation (\ref{eq:velocities-momenta}) between the velocities $\dot{Z} = V$ and $\dot{\Phi} = \Omega$ and the momenta $P = 2\varphi$ and $J = - 2 \zeta$ together with the results for the gyrotropic tensor (\ref{eq:G-nonzero}) yield the internal force and torque: 
\begin{equation}
F_\zeta = 2 \left( \frac{\sin^2{\frac{\varphi}{2}}}{\sinh^2{\frac{\zeta}{2}}} 
	- \frac{\cos^2{\frac{\varphi}{2}}}{\cosh^2{\frac{\zeta}{2}}} \right),
\quad
F_\varphi = - \frac{4\sin{\varphi}}{\sinh{\zeta}}.
\label{eq:forces}
\end{equation}

\begin{figure}
\includegraphics[width=0.99\columnwidth]{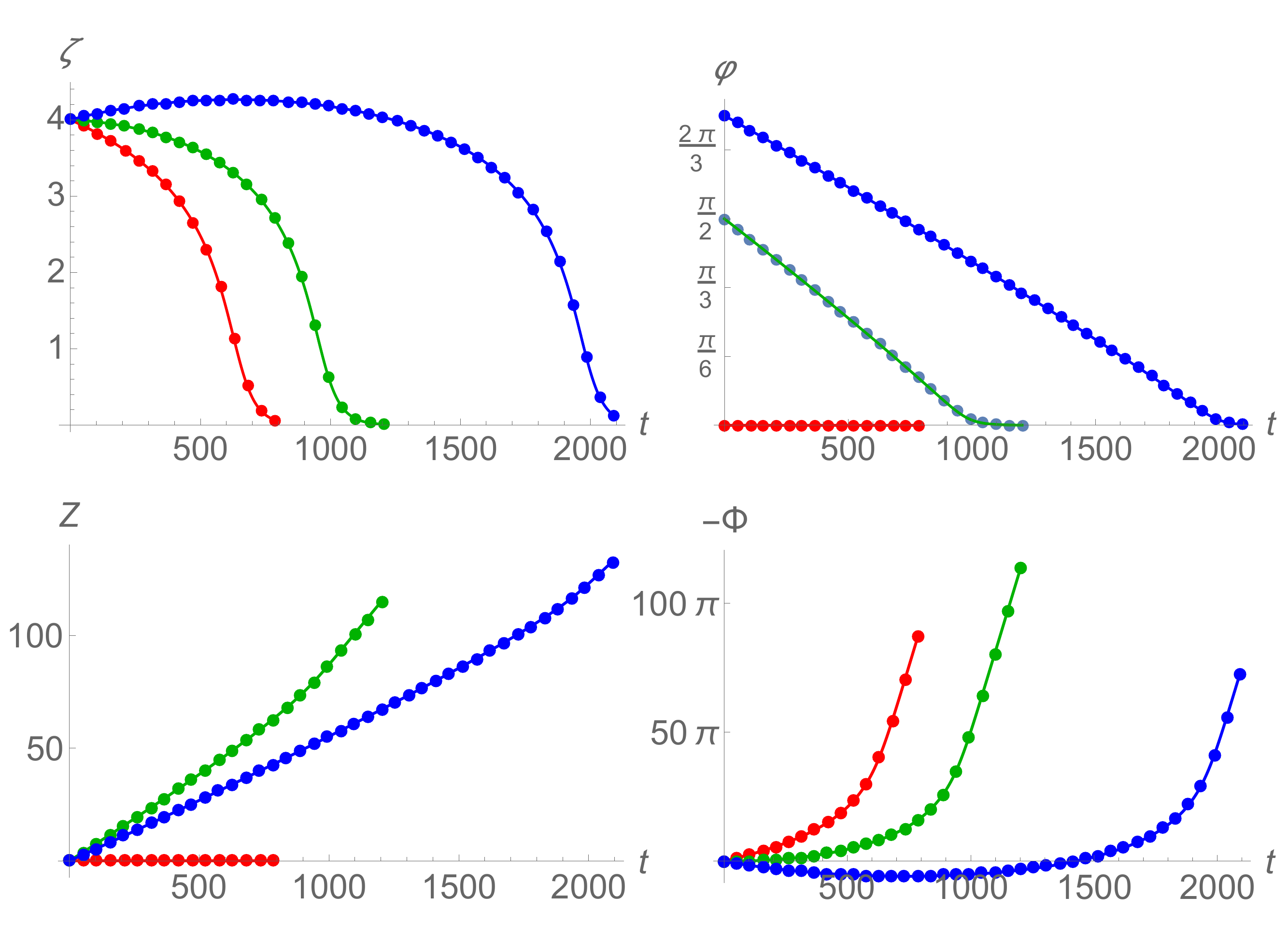}
\caption{(Color online) Collective coordinates $\zeta(t)$, $\varphi(t)$, $Z(t)$, and $\Phi(t)$ for initial separation $\zeta_0 = 4$ and initial twists $\varphi_0 = 0$ (red), $\pi/2$ (green), and $3\pi/4$ (blue). All quantities are in natural units.  Dots are results of micromagnetic simulations, lines are predictions of the effective theory. The Gilbert damping coefficient is $\alpha=0.01$. }
\label{fig:results-plot}
\end{figure}

The viscosity coefficients $\Gamma_{ij}$ are evaluated via Eq.~\ref{eq:q-equations-Gamma} by using the explicit solutions for the solitons \cite{supplemental}. We first focus on the simpler case of zero twist, $\varphi = 0$. In this case, only two collective coordinates, $\Phi$ and $\zeta$, evolve in time, whereas $Z$ and $\varphi$ remain constant. To the lowest non-vanishing order in $\alpha$, 
\begin{equation}
G_{\Phi \zeta} \dot{\zeta} - \Gamma_{\Phi\Phi} \dot{\Phi} = 0,
\quad
F_\zeta + G_{\zeta \Phi} \dot{\Phi} = 0,
\end{equation}
where $F_\zeta = - 2 \sech^2{\frac{\zeta}{2}}$ for $\varphi = 0$. We thus need just one viscosity coefficient, $\Gamma_{\Phi\Phi} = 4 \alpha \tanh{\frac{\zeta}{2}}(1+\frac{\zeta}{\sinh{\zeta}})$ for $\varphi = 0$ \cite{supplemental}. The resulting equations of motion are
\begin{equation}
\dot{\Phi} = F_\zeta/2, 
\quad
\dot{\zeta} = \Gamma_{\Phi\Phi} F_\zeta/4.
\label{eom}
\end{equation}

From Eqs.~\ref{eom} we see that the global rotation angle $\Phi$ is a fast variable whose leading-order behavior is determined by the dissipation-free limit (zeroth order in $\alpha$). Separation $\zeta$ is a slow variable, whose dynamics arises at the first order in $\alpha$ and is dissipational in nature. For large initial separation $\zeta_0 \gg 1$, the attraction is exponentially suppressed, $F_\zeta \approx - 8 e^{-\zeta}$, and the viscosity is approximately constant, $\Gamma_{\Phi \Phi} \approx 4\alpha$. The separation slowly decreases as $\zeta(t) \approx \ln(e^{\zeta_0} - 8 \alpha t)$ until the walls overlap. This initial approach takes an exponentially long time $t_i \approx e^{\zeta_0}/8\alpha$. The final stage, in which the ``separation'' (or angular momentum) decays as $\zeta(t) \sim C e^{-2\alpha t}$, has a characteristic time scale $t_f = 1/2\alpha$. The global rotation frequency initially grows as $\dot{\Phi}(t) \approx -4/(e^{\zeta_0}-8 \alpha t)$ until the walls overlap, then approaches the asymptotic value $\dot{\Phi}_\infty = -1$. 

To check the accuracy of our approach, we compared the solution of Eqs.~\ref{eom} against  numerical simulations of magnetization dynamics in a one-dimensional easy-axis ferromagnet performed with the aid of the micromagnetic solver OOMMF \cite{oommf}. In the simulations, separation $\zeta$ was obtained from the angular momentum along the easy axis, whereas the angle $\Phi$ was measured in the middle of the combined soliton. See Supplemental Material \cite{supplemental} for details. The results for the initial twist $\varphi_0 = 0$, initial separation $\zeta_0 = 4$, and Gilbert damping $\alpha = 0.01$ are shown as red dots (micromagnetic simulations) and red lines (effective theory, Eqs.~\ref{eom}) in Fig.~\ref{fig:results-plot}. We find excellent agreement between the two. 

In the general case, with both an initial twist $\varphi_0 \neq 0$ and separation $\zeta_0 \neq 0$, the equations of motion to the leading order in $\alpha$ have the following form: 
\begin{subequations}
\begin{eqnarray}
\dot{\Phi} = F_\zeta/2,
&\quad&
\dot{\zeta} =  
	(\Gamma_{\Phi \Phi} F_\zeta - \Gamma_{\Phi Z} F_\varphi)/4,
\\
\dot{Z} = -F_\varphi/2,
&\quad&
\dot{\varphi} = 
	(-\Gamma_{Z \Phi} F_\zeta + \Gamma_{Z Z} F_\varphi)/4.
\end{eqnarray}
\label{eom-general}
\end{subequations}
Forces $F_i$ are given in Eq.~\ref{eq:forces}; components of the viscosity tensor $\Gamma_{ij}$ can be found in Supplemental Material \cite{supplemental}. 

During the initial approach ($\zeta \gg 1$), the domain walls interact weakly, $F_\zeta \approx - 8 e^{-\zeta} \cos{\varphi}$, $F_\varphi \approx - 8 e^{-\zeta} \sin{\varphi}$, and retain their individual character, so that the dissipation tensor is diagonal, with $\Gamma_{\Phi\Phi} \approx \Gamma_{ZZ} \approx 4 \alpha$. The twist angle decreases slowly and linearly in time: 
\begin{equation}
\varphi(t) \approx \varphi_0 - 8 \alpha t \, e^{-\zeta_0} \sin{\varphi_0}. 
\end{equation}
The separation evolves as
\begin{equation}
\zeta(t) \approx \zeta_0 + \ln{\frac{\sin{\varphi(t)}}{\sin{\varphi_0}}}.
\end{equation}
Notably, for a large initial twist $\varphi_0 > \pi/2$, the force $F_\zeta$ is repulsive and the domain walls initially move apart until $\varphi$ decreases to $\pi/2$. At that point, the force $F_\zeta$ vanishes and the walls reach their maximum separation $\zeta_\mathrm{max} \approx \zeta_0 - \ln{\sin{\varphi_0}}$. This happens at 
\begin{equation}
t_\mathrm{max} \approx \frac{(\varphi_0 - \pi/2) e^{\zeta_0}}{8 \alpha \sin{\phi_0}}. 
\end{equation}
The total duration of the initial approach is  
\begin{equation}
t_i \approx \frac{\varphi_0 e^{\zeta_0}}{8 \alpha \sin{\varphi_0}}.
\end{equation}
Both the linear trend in $\varphi(t)$ and the backward initial relative motion for $\varphi_0 > \pi/2$ are clearly visible in the micromagnetic simulation data in Fig.~\ref{fig:results-plot}.

During the final stage, the separation and twist decrease to zero. Expanding physical quantities in powers of $\zeta$ and $\varphi$ yields $U \approx 2(\zeta^2 + \varphi^2)/\zeta$, $\Gamma_{\Phi\Phi} \approx 4 \alpha\zeta$, $\Gamma_{Z\Phi} = \Gamma_{\Phi Z} \approx -4 \alpha\varphi$, and $\Gamma_{ZZ} \approx 4 \alpha\varphi^2/\zeta$. Eqs.~\ref{eom-general} read 
\begin{eqnarray}
\dot{\Phi} \approx -1 + \varphi^2/\zeta^2, 
&\quad&
\dot{\zeta} \approx -2 \alpha \zeta (1 + \varphi^2/\zeta^2),
\\
\dot{Z} \approx 2\varphi/\zeta,
&\quad&
\dot{\varphi} \approx -2 \alpha \varphi (1 + \varphi^2/\zeta^2).
\end{eqnarray}
During this stage, the ratio $\varphi/\zeta$ remains constant. Both average velocities attain their terminal values $\dot{Z}_\infty = V_\infty$ and $\dot{\Phi}_\infty = -1 + V_\infty^2/4$, where $V_\infty = 2\varphi/\zeta$. It is interesting to note that, as the domain walls annihilate and the energy decreases toward zero, the pair does not slow down and keeps moving and rotating at constant rates!  The relative coordinates $\zeta(t)$ and $\varphi(t)$ decay exponentially with the characteristic time 
\begin{equation}
t_f \approx \frac{1}{2\alpha(1 + V_\infty^2/4)}. 
\end{equation}
Again, all these trends are clear in Fig.~\ref{fig:results-plot}. The micromagnetic data and the effective theory (Eqs.~\ref{eom-general}) show excellent agreement. 

We have considered the annihilation of two domain walls in a ferromagnetic wire. A minimal description of the process requires the use of 4 physical variables. The average coordinates of the combined soliton, position $Z$ and azimuthal orientation $\Phi$, are zero modes on account of global translational and rotational symmetries; the relative coordinates, separation $\zeta$ and twist $\varphi$, harden as the domain walls merge. We obtained the equations of motion for the these variables in the framework of \textcite{Tretiakov2008} and showed that separation $\zeta$ and twist $\varphi$ exhibit purely viscous dynamics, whereas the average position $Z$ and azimuthal angle $\Phi$ are driven by the torque $F_\varphi(\zeta,\varphi)$ and force $F_\zeta(\zeta,\varphi)$, respectively. These equations of motion (\ref{eom-general}) predict the dynamics of the 4 variables in excellent agreement with the results of numerical micromagnetic simulations (Fig.~\ref{fig:results-plot}). We hope that the method can be successfully extended to the dynamics of other magnetic solitons. 

\emph{Acknowledgments.} Research was supported by the U.S. Department of Energy, Office of Basic Energy Sciences, Division of Materials Sciences and Engineering under Award DE-FG02-08ER46544.

\bibliographystyle{apsrev4-1}
\bibliography{annihilation}

%%%%%%%%%% Merge with supplemental materials %%%%%%%%%%
\pagebreak
\widetext
\begin{center}
\textbf{\large Supplemental Material: Annihilation of Domain Walls in a Ferromagnetic Wire}
\end{center}
\section{Two Domain Wall Soliton}
\noindent
The explicit form of the class of solitons, written as static configurations as seen in the moving frame and parametrized by the four collective coordinates $Z$, $\Phi$, $\zeta$ and $\varphi$, looks like
\begin{subequations}
\label{f_and_g}
\begin{eqnarray}
\cos\theta(z)-1=&f(z-Z;\zeta,\varphi) \label{theta}\\
\phi(z)=&\,\,\,\Phi+g(z-Z;\zeta,\varphi)\label{phi}
\end{eqnarray} 
\end{subequations}
To express the functions $f$ and $g$ in a compact form we first define:
\begin{subequations}
\label{aandb}
\begin{eqnarray}
&a(\varphi,\zeta)=2\Big(1-\cos^2\frac{\varphi}{2}\sech^2\frac{\zeta}{2}\Big) \\ 
&b(\varphi,\zeta)=2\Big(1+\sin^2\frac{\varphi}{2}\csch^2\frac{\zeta}{2}\Big)
\end{eqnarray}
\end{subequations}
The expressions for $f$ and $g$ are then given by:
\begin{subequations}
\label{fandg}
\begin{eqnarray}
&f(z;\zeta,\varphi)=-b+\frac{b-a}{1-\frac{a}{b}\tanh^2\Big(\frac{\sqrt{ab}}{2}z\Big)}\label{f}\\
&g(z;\zeta,\varphi)=\frac{\sin\varphi}{\sinh\zeta}z+\sgn\varphi\tan^{-1}\Bigg[\sqrt{\frac{a(b-2)}{b(2-a)}}\tanh\Big(\frac{\sqrt{ab}}{2}z\Big)\Bigg]\label{g}
\end{eqnarray}
\end{subequations}
Here $\zeta$ and $\varphi$ are independent parameters taking values within the ranges $0<\zeta<\infty$ and $-\pi<\varphi\leq\pi$.

\section{Berry Phase in Effective Lagrangian}
\noindent
Differentiating (4b) in the main text with respect to time gives
\begin{eqnarray}
\dot{\phi}&=\dot{\Phi}+\frac{\partial g}{\partial Z}\dot{Z}+\frac{\partial g}{\partial \zeta}\dot{\zeta}+\frac{\partial g}{\partial \varphi}\dot{\varphi}\nonumber\\
&=\dot{\Phi}-\phi'\dot{Z}+\frac{\partial g}{\partial \zeta}\dot{\zeta}+\frac{\partial g}{\partial \varphi}\dot{\varphi}
\end{eqnarray}
Substituting this into the Berry phase part of the full Lagrangian given by (2) of the main text gives
\begin{eqnarray}
L_{B}&=\dot{\Phi}\displaystyle\int dz\,(\cos\theta-1)-\dot{Z}\int dz\,(\cos\theta-1)\phi'+\dot{\zeta}\int dz\,f\frac{\partial g}{\partial\zeta}+\dot{\varphi}\int dz\,f\frac{\partial g}{\partial\varphi}\nonumber\\
&=-2\zeta\dot{\Phi}+2\varphi\dot{Z}+A_{\zeta}\dot{\zeta}+A_{\varphi}\dot{\varphi}
\end{eqnarray}
The first two terms denote the expected gyrotropic couplings between pairs of cannonically conjugate variables. On the other hand, the gauge connections $A_\zeta(\zeta,\varphi)$ and $A_\varphi(\zeta,\varphi)$ indicate the gyrotropic  coupling between $\zeta$ and $\phi$. The corresponding curvature is
\begin{equation}
F_{\zeta\varphi}=\partial_{\zeta}A_\varphi-\partial_{\varphi}A_\zeta=\int dz\,\Big(\frac{\partial f}{\partial\zeta}\frac{\partial g}{\partial\varphi}-\frac{\partial f}{\partial\varphi}\frac{\partial g}{\partial\zeta}\Big)
\end{equation}
Since $f$($g$) is an even(odd) function of $z$ the integrand is an odd function, implying $F_{\zeta\varphi}=0$. This basically means that under a transport in the $\zeta$-$\varphi$ plane in an infinitesimally closed loop, Berry phases gathered by the spins at equal distances from the center on either side are equal and opposite. This is in agreement with our intuition of what should happen in the large separation limit, when the two domain walls stay rigid under a small transport. Hence the Berry phase part of the effective Lagrangian is
\begin{equation}
L_B=-2\zeta\dot{\Phi}+2\varphi\dot{Z}
\end{equation}

\section{Equations of Motion}
\noindent
Since $F_i=-\frac{\partial U}{\partial q_{i}}$, equations (5a) and (5b) of the main text can be integrated to obtain 
\begin{equation}
U=\frac{4(\cosh\zeta-\cos\varphi)}{\sinh\zeta}
\end{equation}  
The $\Gamma_{ij}(\zeta,\varphi)$ functions form a symmetric $4\times4$ matrix. It is easy to see that this matrix splits up into two $2\times2$ matrices, the $Z$-$\Phi$ block and the $\zeta$-$\varphi$ block, since the remainig four off-diagonal terms are zero. To see this we first express $\Gamma_{ij}$ in terms of $\theta(z)$ and $\phi(z)$.

\begin{equation}
\Gamma_{ij}=\alpha\displaystyle\int dz\,\Big(\frac{\partial\theta}{\partial q_{i}}\frac{\partial\theta}{\partial q_{j}}+\sin^2\theta\frac{\partial\phi}{\partial q_{i}}\frac{\partial\phi}{\partial q_{j}}\Big)
\end{equation}
As seen from (4a-b) of the main text and (\ref{fandg}) above, $\theta(z)$ is even and $\phi(z)$ is odd. Hence if $j=\zeta\,\,\text{or}\,\,\varphi$
\begin{subequations}
\label{gamma_zero}
\begin{eqnarray}
&\Gamma_{Zj}=-\alpha\displaystyle\int dz\,\Big(\frac{\partial\theta}{\partial z}\frac{\partial\theta}{\partial q_{j}}+\sin^2\theta\frac{\partial\phi}{\partial z}\frac{\partial\phi}{\partial q_{j}}\Big)=0 \\
&\Gamma_{\Phi j}=\alpha\displaystyle\int dz\,\sin^2\theta\frac{\partial\phi}{\partial q_{j}}=0
\end{eqnarray}
\end{subequations}
Only the $Z$-$\Phi$ block enters the equations of motion at $O(\alpha)$ and these functions involve only elementary integrals. Performing these integrals, one obtains
\begin{subequations}
\label{gamma_nonzero}
\begin{eqnarray}
&\Gamma_{Z\Phi}=-4\alpha\zeta\frac{\sin\varphi}{\sinh\zeta} \\
&\Gamma_{\Phi\Phi}=\alpha\Big[(b-a)\ln\left\lvert\frac{1+\sqrt{a/b}}{1-\sqrt{a/b}}\right\rvert+2\sqrt{ab}+4(2-b)\tanh^{-1}\sqrt{\frac{a}{b}}\Big]\\
&\Gamma_{ZZ}=2\alpha U-\Gamma_{\Phi\Phi}
\end{eqnarray}
\end{subequations}
The two coupled first-order differential equations for $\zeta(t)$ and $\varphi(t)$ can only be solved numerically. But the large-$t$ limit, when $\zeta,\varphi\to0$ is analytically tractable. In this limit one can expand all the quantities up to the lowest non-vanishing order in the two relative coordinates. This gives
\begin{subequations}
\label{quantities_large t limit}
\begin{eqnarray}
&V=2\varphi/\zeta \,\,\,\,\,\,\,\,\,\,\Omega=(\varphi/\zeta)^2-1 \label{velocities}\\
&U=2(\zeta^2+\varphi^2)/\zeta\\
&\Gamma_{Z\Phi}=-4\alpha\varphi\,\,\,\,\,\,\,\,\,\,\Gamma_{\Phi\Phi}=4\alpha\zeta\,\,\,\,\,\,\,\,\,\,\Gamma_{ZZ}=4\alpha\varphi^2/\zeta
\end{eqnarray}
\end{subequations}
(\ref{velocities}) implies the constraint $\Omega(t)=V(t)^2/4-1$ in this limit. Substituting these expressions in the equations of motion given by (9a-b) of the main text gives
\begin{subequations}
\label{eom_large t limit}
\begin{eqnarray}
&\dot \varphi=-2\alpha(1+\varphi^2/\zeta^2)\varphi \label{eom_1}\\ 
&\dot \zeta=-2\alpha(1+\varphi^2/\zeta^2)\zeta \label{eom_2}
\end{eqnarray}
\end{subequations}
These readily imply $\dot V=(\zeta\dot\varphi-\dot\zeta\varphi)/\zeta^2=0$. Hence both $V$ and $\Omega$ approach constant values $V_{\infty}$ and $\Omega_{\infty}=V_{\infty}^2/4-1$, the exact values of which depend on the choice of initial conditions $\zeta_0$ and $\varphi_0$. This also implies that the two relative coordinates $\zeta$ and $\varphi$ are decoupled in this limit and they both decay exponentially with the same time constant $\tau^{-1}=2\alpha(2+\Omega_{\infty})$.
The shape of the soliton in this limit can be obtained by expanding (\ref{fandg}) to the lowest order terms in the relative coordinates.
\begin{subequations}
\label{fandg_larget}
\begin{eqnarray}
&f(z;\zeta,\varphi)=-\Big(1+\frac{\Omega_{\infty}}{2}\Big)\zeta^2\sech^2\Big[\Big(1+\frac{\Omega_{\infty}}{2}\Big)\zeta z\Big] \\
&g(z;\zeta,\varphi)=\frac{V_\infty}{2}z+\tan^{-1}\Big[\frac{\varphi}{2}\tanh\Big[\Big(1+\frac{\Omega_{\infty}}{2}\Big)\zeta z\Big]\Big]
\end{eqnarray}
\end{subequations}
The expression for $f=m_z(z)-1$ shows that the profile has a characteristic width $w^{-1}=(1+\Omega_{\infty}/2)\zeta$ which diverges as the uniform state is approached.\\\\
\noindent
Animations depicting stationary solitons (in the absence of dissipation) and annihilation processes (in the presence of dissipation) can be found in \href{https://sites.google.com/site/olegtjhu/research-1}{https://sites.google.com/site/olegtjhu/research-1}.
\section{OOMMF Simulation}
\noindent
Micromagnetic simulations of the annihilation process with initial conditions $\zeta_0=4$ (in natural units) and $\varphi_0=0,\pi/2\,\text{and}\,3\pi/4$ were performed using the Object Oriented MicroMagnetic Framework. The parameters used (written as 3-dimensional quantities) are: exchange constant $A=2\times10^{-10}\,\,\text{J}/\text{m}$, anisotropy constant $K=2\times10^4\,\,\text{J}/\text{m}^3$ and magnetization $M_s=10^6\,\,\text{A}/\text{m}$. A wire with cubic unit cell (containing a single spin) of side $a=10\,\,$nm and dimension $N_sa\times a\times a$ was used, $N_s$ being the number of spins in the wire. $N_s$ was taken to be 1000 for $\varphi_0=0$ and 2000 for $\varphi_0=\pi/2\,\,\text{and}\,\,3\pi/4$. The angular momentum density of this system is $\mathcal{J}=M_s/\gamma=5.68\times 10^{-6}\,\,\text{Js}/\text{m}^3$.
This gives a characteristic time scale $t_0=\mathcal{J}/K=0.284\,$ns and a characteristic length scale (exchange length) $\lambda=\sqrt{A/K}=100\,$nm.\\\\
\noindent
Finite-size effects become important toward the end of the simulation for two reasons. Firstly, the width of the soliton diverges as the two domain walls merge, as seen at the end of the previous section. Once it becomes comparable to the system size, the transverse component of magnetization $m_{\perp}$ is no longer zero at the edges of the wire. Secondly, for nonzero $\varphi_{0}$ the soliton has an overall translational motion (in the $+z$ direction for our $\varphi_0$ values), which causes it to run into one of the edges. To tackle the second issue, we started the simulation with $Z_0=-L/4$ for $\varphi_0=\pi/2\,\,\text{and}\,\,3\pi/4$. (Here the origin is at the center of the wire of length $L$.) For $\varphi_0=0$, $Z_0=0$ was used. Moreover, to minimize the finite-size effects, we truncated the simulation data when the value of $m_{\perp}$ at the edge became greater than $1\%$. \\\\
\noindent
We used $\Delta t=0.082\,\text{ns}=0.289t_0$ for each iteration step. The four collective coordinates were extracted from the magnetization profile at each iteration. Since we chose the boundary condition $m_z(z\to\pm\infty)=1$,
\begin{equation}
Z=\frac{ia}{\lambda}\,\,\,\text{where}\,\,\, m_{zi}=\min_{j \in \{1,...,N_s\}}m_{zj}. 
\end{equation}
$\Phi$ is the azimuthal angle of the spin at this location.
\begin{equation}
\cos\Phi=\frac{m_{xi}}{\sqrt{m_{xi}^2+m_{yi}^2}}
\end{equation}
The two relative coordinates are obtained by evaluating the discretized versions of their defining expressions.
\begin{align}
&\zeta=-\frac{a}{2\lambda}\sum_{j=1}^{N_s}(m_{zj}-1)\\
&\varphi=-\frac{1}{2}\sum_{j=1}^{N_s-1}(m_{zj}-1)(\phi_{i+1}-\phi_i)
\end{align}
\end{document}